\documentstyle[12pt,aasms4]{article}
 
\newcommand{\etal}{{\it et al}}

\newcommand{\rne}{\mbox{$N_{\rm e}$}}
\newcommand{\kms}{\mbox{$\rm km\,s^{-1}$}}

\newcommand{\lya}{\mbox{$\rm Ly\alpha$}}
\newcommand{\lyb}{\mbox{$\rm Ly\beta$}}

\lefthead{Sigut & Pradhan}
\righthead{\protect\lya\ Excitation of \protect\ion{Fe}{2}}
 
\begin{document}

\title{Ly$\bf\alpha$ Fluorescent Excitation of Fe\,{\bf\sc ii} in Active Galactic Nuclei}

\author{T.\ A.\ A.\ Sigut}
\affil{Department of Physics and Astronomy,
The University of Western Ontario}
\authoraddr{London, Ontario, Canada N6A 3K7 \\
I: asigut@inverse.astro.uwo.ca}

\and

\author{Anil K. Pradhan}
\affil{Department of Astronomy, The Ohio State University}
\authoraddr{174 West 18th Avenue, Columbus, Ohio 43210-1106 \\
I: pradhan@astronomy.ohio-state.edu}

\slugcomment{To Be Published in The Astrophysical Journal Letters}

\begin{abstract}

We have calculated \ion{Fe}{2} emission line strengths for Active Galactic
Nuclei Broad-Line Regions using precise radiative transfer 
and Iron Project atomic data.
We improve the treatment of all previously considered excitation
mechanisms for the \ion{Fe}{2} emission, continuum fluorescence, collisional
excitation, fluorescence by self-overlap among the iron lines, and fluorescent
excitation by \lya. We demonstrate that \lya\ 
fluorescence is of fundamental importance in determining the strength of
the \ion{Fe}{2} emission. In addition to enhancing the ultraviolet and optical \ion{Fe}{2}
flux, \lya\ fluorescence also results in 
significant near-infrared \ion{Fe}{2} emission in the
$\lambda\lambda8500-9500\,$\AA\ wavelength range. 
New observations are suggested to probe this
effect in strong \ion{Fe}{2} emitting quasars.

\end{abstract}


\section{Introduction}

One of the most puzzling instances of \ion{Fe}{2} line formation
occurs in the spectra of Active Galactic Nuclei (AGN) 
with Broad-Line Regions (BLR)
which exhibit strong ultraviolet \ion{Fe}{2} emission 
(UV$\,\equiv \lambda\lambda2000-3000\,$\AA)
and often strong optical \ion{Fe}{2} emission 
(OP$\,\equiv \lambda\lambda3000-6000\,$\AA) as well.
The widths of the iron lines indicate formation in the
dense BLR, yet standard photoionization calculations, successful in interpreting
many other BLR features, fail to account for the
strength of the \ion{Fe}{2} emission (Wills, Netzer \& Wills 1985; Joly 1993). 
Observed total \ion{Fe}{2}/H$\beta$
ratios are $\approx 12$, whereas
standard models predict ratios less than 8, typically 3 - 5,
with the largest values resulting from enhanced iron abundances.
There is also a class of ``superstrong''
\ion{Fe}{2} emitters with \ion{Fe}{2}/H$\beta \approx 30$ which seem unable to be
explained by traditional photoionization modeling
(Joly 1993; Graham \etal\ 1996).

Excitation mechanisms advanced for the
\ion{Fe}{2} emission include
continuum fluorescence via the UV resonance lines,
self-fluor\-es\-cence via overlapping \ion{Fe}{2} transitions, and
collisional excitation, with the latter thought to contribute the bulk
of the \ion{Fe}{2} emission.
However, Wills \etal\ (1985)
demonstrated that model BLR calculations including these
excitation mechanisms cannot account for
the observed strengths of the \ion{Fe}{2} lines.
Elitzur \& Netzer (1985, hereafter EN) extended this calculation
to include \lya\ fluorescent excitation but found its
contribution to be negligible.
Nevertheless, Penston (1987) argued that
strong indirect observational evidence for the importance of \lya\ fluorescence
is found in the presence of unexpected
UV \ion{Fe}{2} multiplets in the spectrum of the
symbiotic star RR Tel which seem attributable only to cascades from higher
levels pumped by \lya\ fluorescence. The emission nebulae of symbiotics
offer densities and ionization parameters similar to the BLRs of AGN.
Recently, Graham \etal\ (1996) have identified emission from these
UV \ion{Fe}{2} multiplets expected to be preferentially strengthened by \lya\ 
in the spectrum of the ultra-strong \ion{Fe}{2} emitter
2226-3905.

In this {\it Letter}, we improve the modeling of all of the mentioned excitation
mechanisms by using new, accurate atomic data and precise radiative transfer,
and we demonstrate that fluorescent excitation by \lya\ 
can be of prime importance.
A key feature of \lya\ fluorescent excitation of \ion{Fe}{2} is
significant near-IR \ion{Fe}{2} emission
in the wavelength range $\lambda\lambda8500-9500\,$\AA. 
We suggest that a currently well-known, but
unidentified, emission feature near $\lambda9200$ in AGN
spectra may be, at least in part, a spectral signature of this process.

\section{Calculations}

Fe\,{\sc i} - {\sc iv} were considered with
\ion{Fe}{1}, \ion{Fe}{3}, and \ion{Fe}{4} each represented by a
ground state photoionization cross section and total,
unified radiative plus dielectronic recombination
rate from the Iron Project calculations (see Bautista \& Pradhan 1998).
The \ion{Fe}{2} atomic model consists of 262 fine-structure
levels with energies from
Johansson (1978; 1992 private communication).
The model atom is complete
in quartet and sextet levels for energies below 0.849~Rydbergs, allowing
full inclusion of all \lya\ fluorescently excited transitions from $\rm a^4D$.
This atom is a subset of the much larger 827 level \ion{Fe}{2} atom used by
Sigut \& Pradhan (1996).

Transition probabilities for \ion{Fe}{2} were drawn from 3 sources,
Fuhr, Martin \& Wiese (1988), Nahar (1995), and
Kurucz (1991, private communication). 
Over 3,400 \ion{Fe}{2} radiative transitions were included in the
non-LTE solution.
Photoionization cross sections for all \ion{Fe}{2} levels were taken from
Nahar \& Pradhan (1994). A level-specific
radiative plus dielectronic recombination coefficient was used for each
\ion{Fe}{2} quartet and sextet level.
For collisional excitation of \ion{Fe}{2}, we adopted
the \underline{R}-matrix results of Zhang \& Pradhan (1995) and Bautista \&
Pradhan (1998) and used the Gaunt-factor formula for 
all remaining high-lying, dipole transitions.
Charge transfer recombination rates to \ion{Fe}{2} and {\sc iii}
were taken from Kingdon \& Ferland (1996). 
Charge transfer ionization of \ion{Fe}{2}
was also included as \ion{Fe}{3} + $\em e^{-}$ recombination is to the
\ion{Fe}{2} ground state (Neufeld \& Dalgarno 1987).

The coupled equations of radiative transfer and statistical equilibrium
were solved with the accelerated lambda-iteration method of
Rybicki \& Hummer (1992).
As overlap between the \ion{Fe}{2} transitions is
complex, the full preconditioning strategy was implemented.
Complete frequency redistribution (CRD) over
depth-dependent Doppler profiles was assumed for all iron transitions.
The calculations were performed in plane-parallel,
one dimensional, continuum illuminated clouds, each computed for a given
ionization parameter and total particle density, assuming
constant total pressure. The run of
$T_e$ and $\rne$, the photoionizing
radiation field, and the continuous opacities 
and source functions were obtained with
the {\sc cloudy} code (Ferland 1991). This cloud structure was fixed during the
iron non-LTE calculation. The solar iron abundance was 
used, $\log(N_{\rm Fe}/N_{\rm H})+12=7.477$.

The \lya\ source function can be written without 
significant error as
\begin{equation}
S^{\rm{Ly\alpha}}_{\nu}=
\left(\frac{n_{\rm 2p}\,A_{\rm 2p,1s}}{n_{\rm 1s}\,B_{\rm 1s,2p}-
n_{\rm 2p}\,B_{\rm 2p,1s}}\right)\,
\frac{\psi_{\nu}}{\phi_{\nu}}
\label{eq:lyasnu1}
\end{equation}
because stimulated emission is not important.
The quantity in brackets is the usual CRD source function.
The absorption profile,
$\phi_{\nu}$, was taken to be a depth-dependent Voigt profile with a width set
by natural broadening. The emission profile, $\psi_{\nu}$, however,
depends on the radiation field (Mihalas 1978).
We have retained the {\sc cloudy} \ion{H}{1} 1s and 2p level populations,
but have explicitly computed the emission profile in the 
partial coherent scattering approximation
as formulated by Kneer (1975). Assuming a fraction $\gamma_c$ of \lya\ photons
are coherently scattered in the atom's rest frame, 
the ratio of emission to absorption profile is
\begin{equation}
\frac{\psi_{\nu}}{\phi_{\nu}}=1+\gamma_c\left(
\frac{J_{\nu}\cdot\int a_{\nu',\nu}\phi_{\nu'}d\nu'-
\int a_{\nu',\nu}\phi_{\nu'}J_{\nu'} d\nu' }
{\int \phi_{\nu'}J_{\nu'}d\nu'}\right),
\label{eq:lyasnu2}
\end{equation}
where $J_{\nu}$ is the mean intensity, and
$a_{\nu',\nu}$ is a symmetric function that effects the transition from CRD
in the line core to coherent scattering in the wings. We have taken the form of
$a_{\nu',\nu}$ from Vernazza, Avrett \& Loeser (1981)
and adopted $\gamma_c=0.98$. We solved the radiative transfer problem implied by
equation~(\ref{eq:lyasnu2})
separately with an iterative technique, and then the \lya\ 
source function and opacity were incorporated into the appropriate monochromatic
source functions during the solution of the radiative transfer equations for
the \ion{Fe}{2} transitions.
We have verified our solution of equation~(\ref{eq:lyasnu2})
by matching the \lya\ profiles
tabulated for BLR clouds by Avrett \& Loeser (1988) using their models and
hydrogen populations.

\section{Results}

Figure~\ref{fig:lya} compares the predicted \ion{Fe}{2} 
flux with and without \lya\
pumping in BLR models with two different ionization 
parameters ($U_{\rm ion}$), $10^{-2}$ 
and $10^{-3}$. 
These values are chosen to be representative
of typical \ion{Fe}{2}-emitting clouds, but
do not exhaust the range of possibilities.
Both models assumed a total
particle density $4\cdot\,10^{9}\,{\rm cm^{-3}}$, a total column density of
$10^{23}\,{\rm cm^{-2}}$,
and zero microturbulence.
The shape of the photoionizing
continuum was that of Mathews \& Ferland (1987).

For $U_{\rm ion}=10^{-2}$,
the influence of \lya\ is large,
more than doubling the total emitted flux, giving \ion{Fe}{2}/H$\beta = 2.4 $
for UV emission and 0.78 for optical emission, as compared to 1.1 and 0.29,
respectively, without \lya\ 
(the H$\beta$ flux
was taken from the {\sc cloudy} model).
The strongest flux enhancements occur as
noted by Penston (1987). The most important fluorescent excitation rates are from
low-lying $\rm a^4D^e$ levels to the $\rm ^4D^o$, $\rm ^4F^o$, $\rm ^4P^o$,
$\rm ^4S^o$, and $\rm ^6F^o$  symmetries
of the $\rm 3d^6(^5D)5p$ configuration
(Johansson \& Jordan 1984). Cascades 
from these levels to $\rm e^4D$ and $\rm e^6D$ produce 
strong emission lines in the range $\lambda\lambda8500-9500\,$\AA\ 
(see Fig.~\ref{fig:lya2}).
In turn, cascades from $\rm e^4D$ and $\rm e^6D$ 
to odd parity levels near 5~eV
enhance multiplets 399, 391, 380 ($\sim\lambda2850$), 373 ($\sim\lambda2770$),
and 363 ($\sim\lambda2530$); these
lines are the unexpected UV multiplets in the RR Tel spectrum noted by Penston and
the multiplets identified by Graham \etal\ (1996) in 2226-3905. 
Penston estimated that $\approx 20$\%
of the UV \ion{Fe}{2} flux comes out in these multiplets, 
which compares favorably with
the $\approx 30$\% found in the current model.
Subsequent cascades then enhance the
optical and UV fluxes from transitions among the lower levels.
While transitions from $\rm a^4D$ within
$\pm3\,$\AA\ of \lya\
are the most important in the fluorescent process
(see the listing of Johansson \& Jordan 1984),
our treatment of \lya\
includes the entire profile within $\sim \pm50\,$\AA. Including
pumping only within $\pm3\,$\AA\ reduces the flux ratios to 2.1 for UV emission
and 0.70 for optical emission for the case of $U_{\rm ion}=10^{-2}$.

For $U_{\rm ion}=10^{-3}$, the effect of \lya\ pumping
is not as large, particularly on the UV flux.
The \ion{Fe}{2}/H$\beta$ ratio generally increases with decreasing
ionization parameter (even without \lya\ pumping) 
because of the larger sensitivity 
of the hydrogen level populations
to temperature.

As noted by Penston (1987), 
and confirmed by the current calculation (Fig.~\ref{fig:lya2}),
strong observational evidence for \lya\ pumping of \ion{Fe}{2} would
come from the identification of the initial decays from the pumped $\rm 5p$ levels
to $\rm e^4D$ and $\rm e^6D4$ which give rise to transitions in the 
$\lambda\lambda8500-9500\,$\AA\ wavelength range. 
We note that there is a prominent unidentified feature 
near $\lambda9200$ in quasar
spectra as discussed by Morris \& Ward (1989) 
and Osterbrock, Shaw \& Veilleux (1990).
An identification as solely 
\ion{H}{1} Pa9 $\lambda9229$ would render it anomalously
strong compared to Pa8 and Pa10.
Morris \& Ward tentatively assign this feature to \ion{Mg}{2}
$\rm 4p\,^2P^o_{3/2,1/2} - 4s\,^2S_{1/2}$ ($\lambda 9218,9244$)
as formed by cascades from
$\rm 5p\;^2P^o_{3/2,1/2}$ pumped by \lyb. 
They discuss possible contributions
from \ion{Fe}{2} but discount this possibility based on the apparent absence of
a line at $\lambda8927$. Our calculations indicate the $\lambda8927$
\ion{Fe}{2} line is
significantly weaker than the main feature 
near $\lambda9200$ (see Fig.~\ref{fig:lya2}).
We have performed a preliminary non-LTE calculation for
\ion{Mg}{2} including \lyb\ fluorescence in a manner similar to
that described for \lya\ (but with $\gamma_c=0.4$) and find that while
the \ion{Mg}{2} transitions in this region are strengthened by \lyb\ 
by nearly a factor of 5, they are still only minor contributors
to the flux in the $\lambda9200$ region ($<5\,$\%).
Hence, the $\lambda9200$
feature may be, in part, a spectral signature of \lya\ pumping of \ion{Fe}{2}.
As noted by Osterbrock \etal, observations
in this wavelength region are hampered by telluric $\rm H_2O$ absorption bands,
which require careful removal, and, for available observations,
falling CCD responses in the near-IR.
New observations of this wavelength region are warranted.

The current efficiency of \lya\ pumping
is in disagreement with the EN calculation. An
important difference is in 
the adopted oscillator strengths
of the pumped transitions from $\rm a^4D$. The Kurucz (1981)
values are roughly a factor of 10 smaller than those obtained by either
Nahar (1995) or Kurucz (1991, private communication).
Reducing the oscillator strengths for the pumped transitions within
$\pm3\,$\AA\ of \lya\ by a factor of 10
results in a strong reduction in the effect of \lya\ fluorescence, giving for the
$U_{\rm ion}=10^{-2}$
model of Figure~\ref{fig:lya}, \ion{Fe}{2}/H$\beta$ ratios of 1.28 for UV emission
and 0.36 for optical emission. The near-IR emission (Fig.~\ref{fig:lya2}) is also
substantially reduced from \ion{Fe}{2}/H$\beta = 0.326$ to $0.061$. EN also used
a simple escape probability treatment of line overlap, and a simpler
treatment of \lya, so significant differences with the current work are 
perhaps expected.
Finally, we note that our atom is essentially complete in the large number of
quartet and sextet \ion{Fe}{2} energy levels below the $\rm 3d^6(^5D)5p$ configuration,
and that our branching ratios from the pumped levels are expected
to be considerably more accurate than those used by EN.

\section{Conclusions}

Using precise radiative transfer
and accurate atomic data for \ion{Fe}{2},
we demonstrate that
\lya\ fluorescent excitation can more
than double the \ion{Fe}{2} flux in the UV and optical regions. We also predict
\ion{Fe}{2} emission in the $\lambda\lambda8500-9500\,$\AA\ region as a direct result
of \lya\ fluorescence. Both of these results can be viewed as theoretical
confirmation of the role of \lya\ fluorescence suggested by Penston (1987) on the
basis of observations.
New observations of the $\lambda\lambda8500-8500\,$\AA\ wavelength range in
strong \ion{Fe}{2} emitters are highly desirable, and would provide valuable
constraints on modeling $\lya$ fluorescence. Potentially new information
on the \lya\ profile within single BLR clouds could be obtained, something
which is impossible based on observations of the
highly broadened \lya\ profile itself.

A detailed discussion of the variation of the \ion{Fe}{2} 
emission with the physical
parameters defining the BLR clouds (using a much larger 827 level \ion{Fe}{2} atom
which includes the doublet spin system), and the implications for the
`\ion{Fe}{2}/H$\beta$' problem, will be presented by Sigut \& Pradhan (1998).

We are also extending our calculations to overcome some of the limitations of
the present work, in particular, to incorporate all important ions
into a full solution of the non-LTE radiative transfer 
problem for the thermal structure
of AGN BLR clouds. 
This will enable the \lya-\ion{Fe}{2} problem to be solved within
a self-consistent framework.

\acknowledgements
This work was supported by the US NSF through grants to the NCAR
High Altitude Observatory and to AKP (PHY-9421898).
This work was begun while TAAS was an NSERC postdoctoral fellow at HAO.
TAAS wishes to thank J.\ M.\ Marlborough and
J.\ D.\ Landstreet for support through their NSERC grants.

%
%

\newpage

\figcaption[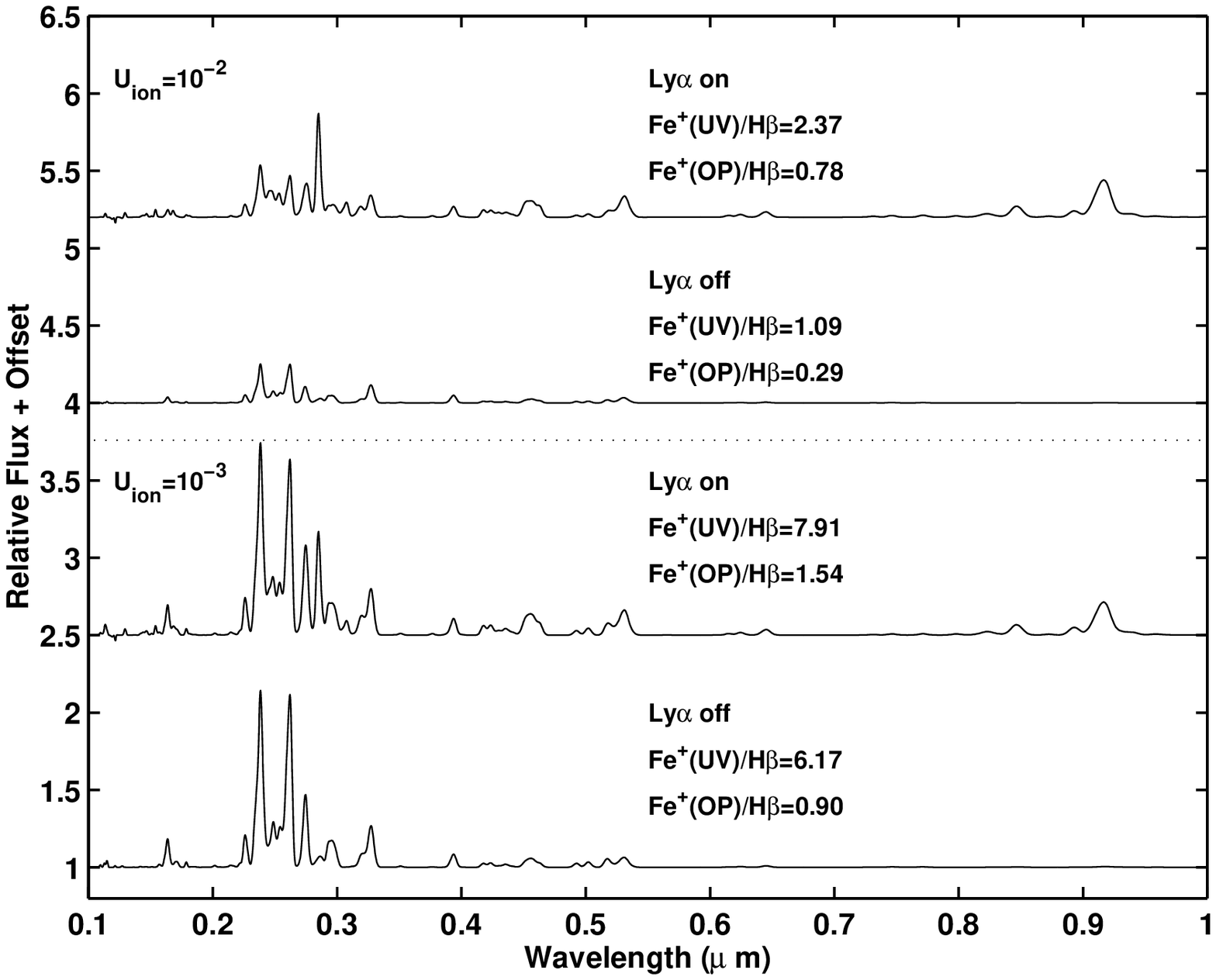]{The predicted \protect\ion{Fe}{2} spectrum.
The spectra were convolved with a Gaussian of $1/e$ width of 2000~\protect\kms.
Covering fractions of
the illuminated and shielded cloud faces were 
assumed to be 5\% each. The ratios of
\protect\ion{Fe}{2} fluxes with H$\beta$ are computed with the optical 
(OP) and ultraviolet (UV)
wavelength regions defined in the introduction.
\label{fig:lya}}

\figcaption[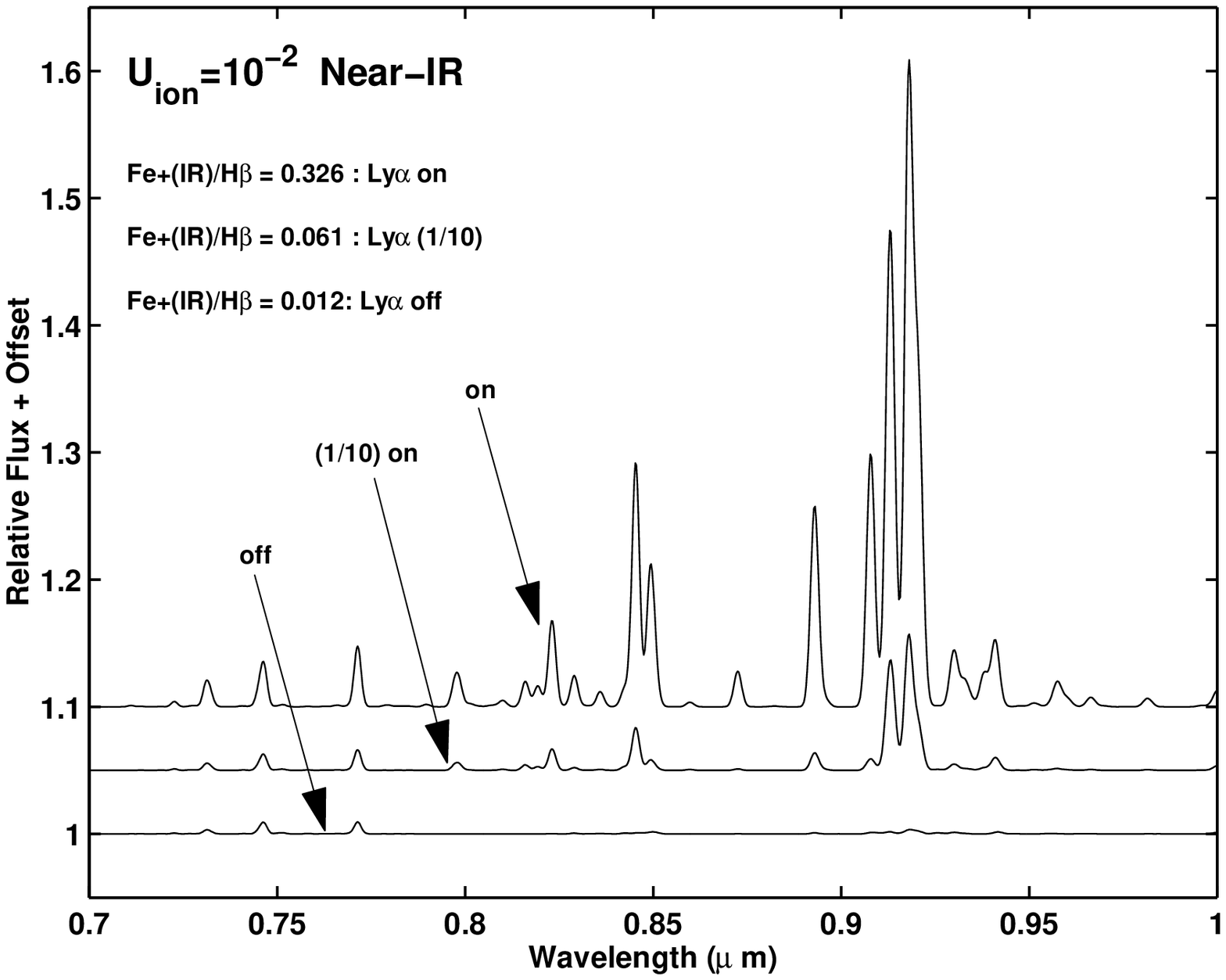]{An enlargement of the near-IR \protect\ion{Fe}{2} 
spectrum for the $U_{\rm ion}=10^{-2}$
model of Figure~1.
The spectra were convolved with a Gaussian of $1/e$ width of
500~\protect\kms. `\lya\ (1/10)' indicates that the $\rm A_{ji}$
values of the pumped \protect\ion{Fe}{2} 
transitions within $\pm 3\,$\AA\ of \lya\ 
were scaled by $0.1$. \protect\ion{Fe}{2}(IR) 
refers to the flux between $0.7 - 1.0\,$\micron.
\label{fig:lya2}}


\begin{references}

\reference{} Avrett, E.\ H., \& Loeser, R.\ 1988 \apj\ 331, 211

\reference{} Bautista, M.\ A., \& Pradhan, A.\ K.\ 1998 \apj\ 492, 650

\reference{} Elitzur, M., \& Netzer, H.\ 1985 \apj\ 291, 464

\reference{} Ferland, G.\ J.\ 1991 {\it Hazy: An Introduction to Cloudy},
OSU Internal Report 91-01

\reference{} Fuhr, J.\ R., Martin, G.\ A., \& Wiese, W.\ L.\ 1988 J.\ Phys.\
Chem.\ Ref.\ Data 17, 1

\reference{} Graham, M.\ J., Clowes, R.\ G., \& Campusano, L.\ E.\ 1996 
\mnras\ 279, 1349

\reference{} Johansson, S., 1978 Physica Script 18, 217

\reference{} Johansson, S., Jordan, C.\ 1984 \mnras\ 210, 239

\reference{} Joly, M.\ 1993 Ann.\ Phys.\ Fr.\ 18, 241

\reference{} Kingdon, J.\ B., \& Ferland, G.\ J.\ 1996 \apjs\ 106, 205

\reference{} Kneer, F.\ \apj\ 200, 367

\reference{} Kurucz, R.\ L.\ 1981, {\it Semi-empirical Calculation of gf values :
\protect\ion{Fe}{2}}, Smithsonian Astrophysical Observatory Report 390 (Cambridge)

\reference{} Mathews, W.\ G., \& Ferland, G.\ J.\ 1987 \apj\ 323, 456

\reference{} Mihalas, D.\ 1978 {\it Stellar Atmospheres}, 2nd edition
(W.\ H.\ Freeman \& Company : San Fransico)

\reference{} Morris, S.\ L., \& Ward, M.\ J.\ 1989 \apj\ 340, 713

\reference{} Nahar, S.\ N.\ 1995 \aap\ 293, 967

\reference{} Nahar, S.\ N., \& Pradhan A.\ K.\ 1994 
J.\ Phys.\ B: At.\ Mol.\ Opt.\ Phys.\ 27, 429

\reference{} Neufeld, D.\ A., \& Dalgarno, A.\ 1987 \pra\ 35, 3142

\reference{} Osterbrock, D.\ E., Shaw, R.\ A., \& Veilleux, S.\ 1990 
\apj\ 352, 561

\reference{} Penston, M.\ V.\ 1987 \mnras\ 229, 1p

\reference{} Rybicki, G.\ B., \& Hummer, D.\ G.\ 1992 \aap\ 262, 209

\reference{} Sigut, T.\ A.\ A., \& Pradhan, A.\ K.\ 1996 \baas\ 28, 1405

\reference{} Sigut, T.\ A.\ A., \& Pradhan, A.\ K.\ 1998 \apj\ to be submitted

\reference{} Vernazza, J.\ E., Avrett, E.\ H., \& Loeser, R.\ 1981 \apjs\ 45, 635

\reference{} Wills, B.\ J., Netzer, H., \& Wills, D.\ 1985 \apj\ 288, 94

\reference{} Zhang, H.\ L., \& Pradhan, A.\ K.\ 1995 \aap\ 293, 953

\end{references}
\end{document}